\documentclass[12pt]{article}
\usepackage{amsfonts}
\usepackage{epic}
\newcommand{\si}{\sigma}

\newcommand{\nn}{\nonumber}

\newcommand{\be}{\begin{equation}}
\newcommand{\ee}{\end{equation}}
\newcommand{\bea}{\begin{eqnarray}}
\newcommand{\ena}{\end{eqnarray}}
\newcommand{\beas}{\begin{eqnarray*}}
\newcommand{\enas}{\end{eqnarray*}}

\def\ZZ{{\mathbb Z}}

\begin{document}
\pagestyle{empty}

\begin{center} 
{\Large  {Staggered Anisotropy Parameter Modification of                   
      \\[4mm]
      the Anisotropic $t-J$ Model}} 
  
\vspace{36pt}

{\bf T.~Sedrakyan\footnote{e-mail:{\sl tigrans@moon.yerphi.am,
Permanent address: Yerevan Physics Institute, Armenia}}}\\

\vfill

\emph{Laboratoire d'Annecy-le-Vieux de Physique Th{\'e}orique LAPTH}
\\
\emph{CNRS, UMR 5108, associ{\'e}e {\`a} l'Universit{\'e} de Savoie}
\\
\emph{BP 110, F-74941 Annecy-le-Vieux Cedex, France}
\\

\vfill
{\bf Abstract}

\end{center}

The anisotropic t-J model ($U_q(gl(2|1))$ Perk-Schultz model) with
staggered disposition of the anisotropy parameter along a chain
is considered and the corresponding ladder type integrable model is
constructed. This
is a generalisation to spin-1 case of the staggered $XXZ$ spin-1/2 model
considered earlier. The corresponding Hamiltonian
is calculated and, since it contains next to nearest neighbour
interaction terms, can be written in a zig-zag form. The Algebraic
Bethe Ansatz technique is applied and the eigenstates, along with
eigenvalues of the transfer matrix of the model are found.

\vfill
\rightline{LAPTH-840/01}
\rightline{hep-th/0103027}
\rightline{February 2001}

\newpage
\pagestyle{plain}
\setcounter{page}{1}

\section{Introduction}
\indent

The interest to ladder type models was raised in a beginning of
90-s (see  \cite{Rice} for a review) in connection with high
temperature superconductivity problems in metal oxides. It is
believed that quasi-one dimensional multi-ladder chains of
strongly interacting electrons reflect the most important
aspects of two dimensional systems and also can reveal some 
properties of the weak coupling between conducting planes.

Recently there has been considerable interest in the construction
of integrable ladder type models motivated by the desire to use
powerful technique of Algebraic Bethe Ansatz (ABA) \cite{Bax,FT} 
in the exact investigations of the variety of physical phases
of the models.

In the articles \cite{Wa} integrable ladder models were
constructed by extension of the symmetry algebra, in \cite{Ko}
by defining first the ground state and then formulating a model,
which has it. The higher conservation laws of an integrable models,
which contains next to nearest neighbour interactions, were
used in construction of ladder models in \cite{Muu} by developing
the approach of the article \cite{Maj}.The models with
alternating spins were considered in \cite{VMN,DM}.
There are also some other  attempts in this area \cite{Al,For}.

Usually integrable models are homogeneous along the chain, namely,
the spectral $u$ and model parameters are the same in the
product of $R$-matrices along the chain. It is obvious, that if one 
considers arbitrary shifts of the spectral parameters by
some $z_i$ in the monodromy matrix we still have an integrable
model. But in order to have a local Hamiltonian we need to consider
shifts with fixed periodicity $n$, which causes the
interaction of spins (or electrons) within an amount of $n$ neighbours,
leading to $n$-ladder model. The staggered shift of the spectral parameter
was first considered in \cite{Destri} in an attempt to construct a
relativistic invariant massive Tirring Model in a specific limit
of the homogeneous $XXZ$-model. 

In recent articles \cite{APSS, AASSS} the authors have proposed
a construction of some integrable chain
models with $\mathbb{Z}_2$ grading of the
states along the chain, as well as along the time directions.
Hence, a way of constructing an integrable models
with staggered inhomogeneity was proposed, motivated
by the problems of three dimensional Ising model \cite{KS}
and the Hall effect \cite{S}. The inhomogeneous anisotropic
$XXZ$ Heisenberg chain with staggered anisotropy
parameter $\pm \Delta$ was constructed in \cite{APSS},
while the isotropic $t-J$ model was considered in \cite{AASSS}.

The $XXZ$ and $t-J$ models with only inhomogeneous shift of
the spectral parameter was previously considered in a chain
of articles \cite{FLec, ZV, ZV1, FR}, but these authors have not 
analysed the possibility of the inhomogeneity of the anisotropy
parameter $\Delta$ (or other model parameters). As a result they
have the same intertwining $R$-matrix as for ordinary
homogeneous integrable $XXZ$  and $t-J$ models correspondingly
and, therefore, the same quantum group structure behind.
Contrary to this, in the construction presented in the
articles \cite{APSS} and \cite{ASSS} the integrable inhomogeneity
appeared not only in shifts of the spectral parameter, but
also in some structural changes. As a result we got modified
Yang--Baxter equations (YBE), leading to 
a quantum group algebraic structure different from the usual $sl_q(2)$
\cite{ASSS}. 

As it is shown in the mentioned above articles, due to periodic
shift of the spectral parameters the Hamiltonians of all this
models contain at most next to nearest neighbour (NNN) interaction
terms and therefore can be represented as  integrable two
leg ladder (zig-zag) models. The interaction between the legs
of the ladder is represented in the Hamiltonian by a topological
type terms written as  the product of three neighbour spins and the
anisotropic antisymmetric tensor as
\be
\label{0}
\hat{\epsilon}^{abc}\sigma_i^a \sigma_{i+1}^b  \sigma_{i+2}^c.
\ee

As it was shown in \cite{S}, the model of free fermions hopping
with 
inhomogeneous parameters, in a case when the rotational
invariance is preserved, determines a Peierls type mechanism
of mass generation, which is based on the breaking of the 
translational invariance for the translations of one lattice
spacing. Perhaps the above mentioned 3-spin
interaction terms are responsible for a gaped phase of the
system, as it is discussed in the articles \cite{Wen} in connection
with high temperature superconductivity problems.

In this article, by use of the solution of the staggered YBE
for the general $Sl_q(n)$ quantum group presented in \cite{ASSS},
we pass from spin-1/2 $XXZ$ model to three state $gl(2|1)$ supergroup based
anisotropic $t-J$ model \cite{PS, dV, ACF, FK} (or $U_q(gl(2|1))$ 
Perk-Schultz model)
with staggered sign of the anisotropy parameter. Following the technique
of fermionization of spin models developed in \cite{AK} 
one can see, that anisotropic supersymmetric $t-J$ model cam be regarded
as fermionized version of the anisotropic $spin 1$ bosonic
Uimin-Lai-Sutherland model \cite{U,L,Su}.

In Section 2 we present the solution of the
staggered $YBE$ for the $Sl_q(1,2)$ case and calculate the
Hamiltonian of the model in a ladder form.

In Section 3 we apply the technique of the Algebraic
Bethe Ansatz $(ABA)$ \cite{Bax,FT} and find the eigenvalues and 
eigenstates
of our Hamiltonian. The Bethe equations $(BE)$ for the 
allowed values of the spectral parameters (momentums) are
written down here.


\section{The Yang--Baxter Equations, their solution and the Hamiltonian
  of the Model}
\indent

In this Section we follow the technique developed in the articles
\cite{APSS, AASSS}. We modify the anisotropic supersymmetric $t-J$ 
model (which is the $U_q(gl(2|1))$ Perk-Schultz model \cite{PS, dV, ACF})
 in order to construct an integrable
model with staggered disposition of the sign of the 
anisotropy parameter $\Delta$.

The principal ingredient of integrable models via Bethe Ansatz
technique is the $R$-matrix. The $R_{aj}$-matrix acts as an intertwining
operator on the direct product of the
auxiliary $V_a(v)$ and quantum $V_j(u)$ spaces 
\begin{eqnarray}
\label{5}
R_{aj}(u,v): V_a(u)\otimes V_j(v)\longrightarrow V_j(v)\otimes V_a(u) \;.
\end{eqnarray}

For the anisotropic $t-J$ model the quantum and auxiliary
spaces are three dimensional and 
corresponding $R$-matrix (see for example \cite{dV}) can be 
defined by the following formula
\begin{eqnarray}
\label{3}
R_{aj}(z)=qzR_{aj,q}-q^{-1}z^{-1}R_{ja,q}^{-1},
\end{eqnarray}
where we have introduced a multiplicative spectral parameter
$z=e^{iu}$. The parameter $q$ defines the anisotropy of the model and, 
as it is shown in \cite{ACF}, the so called 
constant $R$-matrix $R_q$ is equal to
\begin{eqnarray}
\label{4}
R_{aj,q}=\lim_{z\rightarrow\infty}\frac{R(z)}{z} =\left( 
\begin{array}{lllllllll}
-1& 0& 0& 0& 0& 0& 0& 0& 0\\
0&-q^{-1}& 0& 0& 0& 0& 0& 0& 0\\
0& 0& q^{-1}& 0& 0& 0& 0& 0& 0\\
0& -q^{-1}\lambda& 0& -q^{-1}& 0& 0& 0& 0& 0\\
0& 0& 0& 0& -1& 0& 0& 0& 0\\
0& 0& 0& 0& 0& q^{-1}& 0& 0& 0\\
0& 0& -q^{-1}\lambda& 0& 0& 0& q^{-1}& 0& 0\\
0& 0& 0& 0& 0& -q^{-1}\lambda& 0& q^{-1}& 0\\
0& 0& 0& 0& 0& 0& 0& 0& q^{-2}
\end{array}\right),
\end{eqnarray}
with $\lambda = q-q^{-1}$.

It is very convenient for this model to consider its fermionized
version by use of the technique developed in the article \cite{HS, AK}.
Let us introduce two Fermi fields $c_{i,\sigma},\; c^+_{i,\sigma}$
with spin up and down states $\sigma =\uparrow, \downarrow$
at each site $i$ of the chain and  three basic vectors of the
corresponding spaces $V_i$ as follows
\begin{equation}
\label{6}
\mid-\rangle \equiv \mid0 ,\downarrow \rangle, \qquad
\mid+\rangle \equiv \mid \uparrow ,0 \rangle,  \qquad
\mid0\rangle \equiv \mid0 ,0\rangle,
\end{equation}
numerated as $\mid 1 \rangle,\mid 2 \rangle,\mid 3 \rangle$
respectively. Therefore we have $\mathbb{Z}_2$ graded quantum $V_j(u)$
and auxiliary $V_a(v)$ spaces with the following parities
\begin{equation}
\label{7}
p(\mid+\rangle)=p(\mid -\rangle)=1, \qquad p(\mid 0\rangle)=0.
\end{equation}

Consider now Hubbard operators 
${X}_{a_2}^{a_1}=\mid a_2\rangle\langle a_1\mid,\; 
{X}_{j_2}^{j_1}=\mid j_2 \rangle \langle j_1\mid$, with $\mid a\rangle,
\mid j\rangle$ defined as in formulas (\ref{6}) and
\begin{eqnarray}
\label{8}
X_{ m}^k &=&\left( 
\begin{array}{lll}
|-\rangle \langle -|&\qquad |- \rangle \langle +| 
&\qquad |-\rangle \langle 0| \\
|+\rangle \langle -|&\qquad |+\rangle \langle +| 
&\qquad |+\rangle \langle 0|\\ 
|0\rangle \langle -|&\qquad |0\rangle \langle +| 
&\qquad |0\rangle \langle 0|  
\end{array}
\right) \nonumber\\
&=&\left( 
\begin{array}{lll}
(1-n_{\uparrow}) n_{\downarrow}&c^{+}_{\downarrow} c_{\uparrow}
 &(1-n_{\uparrow})c^{+}_{\downarrow} \\
c^{+}_{\uparrow} c_{\downarrow}& n_{\uparrow}(1- n_{\downarrow})
& c^{+}_{\uparrow}(1-n_{\downarrow}) \\ 
(1-n_{\uparrow})c_{\downarrow} & c_{\uparrow}(1-n_{\downarrow})
&(1-n_{\uparrow})(1- n_{\downarrow}) 
\end{array}
\right),
\end{eqnarray}  
where $n_{\sigma}=c^{+}_{\sigma}c_{\sigma}$ is the particle number
operator. The demand that the trace of this operator acts on auxiliary and
quantum spaces as the identity operator means that the double
occupancy of the sites by fermions is excluded
\begin{equation}
\label{9}
\Delta \mid l \rangle =X_m^m \mid l \rangle =
(1-n_{\uparrow}n_{\downarrow}) \mid l \rangle =\mid l \rangle,\qquad
m,l=1,2,3. 
\end{equation}

Now we can write down the formula connecting the fermionic 
$R_{aj}$-operator of the anisotropic supersymmetric $T-J$ model
with the $R_{aj}$-matrix
defined by the expression-s (\ref{3}) and (\ref{4}) 
\begin{eqnarray}
\label{10}
R_{aj}&=&R_{aj}\mid {j_1}\rangle\mid {a_1}\rangle \langle {a_1}
\mid\langle {j_1}\mid = 
(-1)^{p(a_1)p(j_2)}(R_{aj})^{a_2j_2}_{a_1j_1}X^{a_1}_{a_2}X^{j_1}_{j_2}.
\end{eqnarray}
It is straightforward to check that this fermionic $R$-operator
satisfies the YBE in the operator form \cite{AK}.

Let us now consider $\ZZ_2$ graded quantum $V_{j,\rho}(v)$ and 
auxiliary $V_{a,\si}(u)$ spaces, $\rho, \si =0,1$. In this case 
we  have $4\times 4$ $R$-matrices, which act on the direct product
of the spaces $V_{a,\si}(u)$ and $ V_{j,\rho}(v)$, $(\si,\rho =0,1)$,
mapping them on the intertwined direct product of 
$V_{a,\bar{\si}}(u)$ and $ V_{j,\bar{\rho}(v)}$ with the complementary
$\bar{\si}=(1-\si)$, $\bar{\rho}=(1-\rho)$ indices
\begin{equation}
\label{1R5}
R_{aj,\si \rho}\left( u,v\right):\quad V_{a,\si}(u)\otimes 
V_{j,\rho}(v)\rightarrow V_{j,\bar{\rho}}(v)\otimes V_{a,\bar{\si}}(u).  
\end{equation}

It is convenient to introduce two transmutation operations $\iota_1$
and $\iota_2$ with the property $\iota_1^2=\iota_2^2=id$ 
for the quantum and auxiliary spaces
correspondingly, and to mark the operators $R_{aj,\si\rho}$ as 
follows
\begin{eqnarray}
\label{R51}
R_{aj,00}&\equiv& R_{aj},\qquad R_{aj,01}\equiv R_{aj}^{\iota_1},\nn\\
R_{aj,10}&\equiv& R_{aj}^{\iota_2},\qquad R_{aj,11}\equiv R_{aj}^{\iota_1 
\iota_2}.
\end{eqnarray}

The introduction of the $\ZZ_2$ grading in quantum space means, 
that we have now two monodromy matrices $M_{\rho}, \rho=0,1$,
which act on the space $V_{\rho}=\prod_{j=1}^N V_{j,\rho}$
by mapping it on $V_{\bar{\rho}}=\prod_{j=1}^N V_{j,\bar{\rho}}$
\begin{equation}
\label{1M2}
M_\rho \qquad : V_\rho \rightarrow V_{\bar{\rho}}, \qquad \qquad 
\rho=0,1.
\end{equation}

It is clear now, that the monodromy matrix of the model, which should 
define the partition function, is the product of two monodromy matrices
\begin{equation}
\label{1}
M(u) = M_0(u) M_1(u).
\end{equation}

Now, because of the grading in the auxiliary space, we would like to
construct the monodromy matrices $M_{0,1}$ as a staggered product
of the $R_{aj}$ and $\bar{R}_{aj}^{\iota_2}$ matrices.
Let us define
\begin{eqnarray}
\label{2}
M_0(u)=\prod_{j=1}^N R_{a,2j-1}(u)
\bar{R}_{a,2j}^{\iota_2}(u)\nn\\
M_1(u)=\prod_{j=1}^N \bar{R}_{a,2j-1}^{\iota_1}(u)
R_{a,2j}^{\iota_1 \iota_2}(u),
\end{eqnarray}
where the notation $\bar{R}$ in general means the
different parametrisation of the $R$-matrix via model
($\lambda$) and spectral ($u$) parameters and can be considered as an
operation over $R$ with property $\bar{\bar{R}}= R$.

In order to have a integrable model with commuting transfer matrices
\begin{equation}
\label{11}
\tau(u)=tr_0 tr_{0^{\prime}} M(u) = \tau_0(u)\tau_1(u).
\end{equation}
for different spectral parameters
\begin{equation}
\label{T5}
\left[ \tau(u), \; \tau(v)\right]=0
\end{equation}
it is enough to have the following relations for the $\tau_{\si}(u)=
trM_{\si}(u), \qquad (\si=0,1)$
\begin{equation}
\label{T51}
\tau_{\si}(\lambda,u)\; \tau_{1-\si}(\lambda,v)=
\bar{\tau}_{\si}(\lambda,v)\; \bar{\tau}_{1-\si}(\lambda,u),\qquad \si=0,1
\end{equation}

It is not hard to see, that in order to ensure the commutativity
condition (\ref{T5}) the $R$- and $\bar{R}$-matrices in the expression 
(\ref{2}) should fulfill the following two Yang-Baxter
Equations, which in so called check formalism defined by operator
$\check{R}_{ij}=R_{ij} P_{ij}$, ($P_{ij}$ is the permutation
operator), has the form
\begin{equation}
\label{12.1}
\check{R}_{12}(u,v)\check{\bar{R}}_{23}^{\iota_1}(u)\check{R}_{12}(v)=
\check{R}_{23}^{\iota_1}(v)\check{\bar{R}}_{12}(u)\check{\tilde{R}}_{23}(u,v),
\end{equation}
and
\begin{equation}
\label{12.2}
\check{\tilde{R}}_{12}(u,v)\check{R}_{23}^{\iota_1\iota_2}(u)
\check{\bar{R}}_{12}^{\iota_2}(v)
=\check{\bar{R}}_{23}^{\iota_1\iota_2}(v)\check{R}_{12}^{\iota_2}(u)
\check{R}_{23}(u,v).
\end{equation}

If the $\iota_1$ and $\iota_2$ operations are trivial this $YBE$'s are
reducing to ordinary ones. Therefore one can take an integrable model
with the $R$-matrix fulfilling $YBE$ and try to find nontrivial
$\iota_1$ and $\iota_2$ operations. For the $Sl_q(N)$ case this type 
of solutions were found in \cite{ASSS} from where we will extract the
solution for the $Sl_q(2|1)$ case under consideration. The solution
of the first set of equations (\ref{12.1}) with the nontrivial
$\iota_1$ operation looks like

\begin{eqnarray}
\label{13}
\check{R}^{\iota_1}(z)=qz\check{R}^{\iota_1}_q-
q^{-1}z^{-1}(\check{R}^{\iota_1}_q)^{-1}
\end{eqnarray} 
with
\begin{eqnarray}
\label{14}
\check{R}^{\iota_1}_q=\tilde{R}_q
 =\left( 
\begin{array}{lllllllll}
-1& 0& 0& 0& 0& 0& 0& 0& 0\\
0& -q^{-1}\lambda & 0& -q^{-1}& 0& 0& 0& 0& 0\\
0& 0& -q^{-1}\lambda & 0& 0& 0& -q^{-1}& 0& 0\\
0& -q^{-1} & 0& 0 & 0& 0& 0& 0& 0\\
0& 0& 0& 0& -1 & 0& 0& 0& 0\\
0& 0& 0& 0& 0& -q^{-1}\lambda & 0& -q^{-1}& 0\\
0& 0& -q^{-1} & 0& 0& 0& 0 & 0& 0\\
0& 0& 0& 0& 0& -q^{-1} & 0& 0& 0\\
0& 0& 0& 0& 0& 0& 0& 0& q^{-2}
\end{array}\right).
\end{eqnarray}
and the bar operation in (\ref{12.1}-\ref{12.2}) is a multiplicative 
shift of the multiplicative
spectral parameter $z=e^{iu}$ by the new model parameter $h=e^{i \theta}$
as $\check{R}(\bar{z})=\check{R}(e^{i\bar{u}})= \check{R}(hz^{-1})$. 

This definition for the $\iota_1$ operation means that
\be
\label{RI}
(R_q^{\iota_1})_{13}^{13}=-(R_q)_{13}^{13},\;\;
(R_q^{\iota_1})_{23}^{23}=-(R_q)_{23}^{23},\;\;
(R_q^{\iota_1})_{12}^{12}=-(R_q)_{12}^{12}.
\ee

If we define the operation $\iota _2$ on $\check{R}(z)$ as
\begin{equation}
\label{15}
\check{R}^{\iota _2}(z)=\check{R}^{\iota _1}(z^{-1}),
\end{equation}
then easy to see that second set of YBEs will coincide with the first
set of $YBE$-s.

The explicit expressions for the matrix elements of $R(z)$ and
$R^{\iota_1}(z)$ matrices which are fulfilling the $YBE$-s 
(\ref{12.1}-\ref{12.2}) are the followings
\bea
\label{exR1}
R_{11}^{11}(z)&=&R_{22}^{22}(z)=q^{-1}z^{-1}-q z,\;\;\;\;\;\;
R_{33}^{33}(z)=q^{-1}z-q z^{-1},\nn\\
R_{11}^{33}(z)&=&R_{22}^{33}(z)=R_{33}^{11}(z)=R_{33}^{22}(z)=
-R_{11}^{22}(z)=-R_{22}^{11}(z)=z-z^{-1}\nn\\
R_{21}^{12}(z)&=&R_{31}^{13}(z)=R_{32}^{23}(z)=-z\lambda,\;\;
R_{12}^{21}(z)=R_{13}^{31}(z)=R_{23}^{32}(z)=-z^{-1}\lambda
\ena
and
\bea
\label{exR2}
(R^{\iota_1}(z))_{11}^{11}&=&(R^{\iota_1}(z))_{22}^{22}=q^{-1}z^{-1}-q z,\;\;\;\;\;\;
(R^{\iota_1}(z))_{33}^{33}=q^{-1}z-q z^{-1},\nn\\
(R^{\iota_1}(z))_{11}^{33}&=&(R^{\iota_1}(z))_{22}^{33}=(R^{\iota_1}(z))_{33}^{11}=
(R^{\iota_1}(z))_{33}^{22}=
(R^{\iota_1}(z))_{11}^{22}=(R^{\iota_1}(z))_{22}^{11}=-z+z^{-1}\nn\\
(R^{\iota_1}(z))_{21}^{12}&=&(R^{\iota_1}(z))_{31}^{13}=
(R^{\iota_1}(z))_{32}^{23}=-z\lambda,\nn\\
(R^{\iota_1}(z))_{12}^{21}&=&(R^{\iota_1}(z))_{13}^{31}=
(R^{\iota_1}(z))_{23}^{32}=-z^{-1}\lambda
\ena

Therefore we have found an integrable inhomogeneous model and
our aim now is to write down the expression for the Hamiltonian.
But in order to have a local Hamiltonian it is
necessary to have a point $u_0$, where
\begin{equation}
\label{16}
\check{R}_{ij}(u_0)= const. \hat{I} _{ij},
\end{equation}
with $\hat{I} _{ij}$ being an identity operator.

>From the form of $\check{R}$-matrix (\ref{13}) one can easily see
that the value $u_0=0$ (or $z_0=e^{iu_0}=1)$ is just the needed one. Let us
notice that though $\check{R}_{ij}(u)\arrowvert _{u=0}=-\lambda \hat{I}_{ij}$,
we have   
$\check{R}_{ij}(\bar u)\arrowvert _{u=0} \neq \hat{I} _{ij}$.
Because of this, and as
future calculations will show, the Hamiltonian of our model contains, 
together with
nearest-neighbour interaction terms, also next-to
nearest-neighbour interaction terms.

By definition the Hamiltonian of the model is the logarithmic
derivative of the transfer matrix at that point
\begin{eqnarray}
\label{17}
\mathfrak{H}=-\frac{\partial {\ln{\tau(u)}}}{\partial{u}}{\arrowvert}_{u=0}.
\end{eqnarray} 

For calculations of logarithmic derivative of the Transfer
matrix (\ref{11}) we need to insert the 
linear expansions of the operators $\check{R}(z), 
\check{R}^{\iota _1\iota _2}(z), \check{R}^{\iota _1}(\bar z), 
\check{R}^{\iota _2}(\bar z)$ around the point $u=0$
\begin{eqnarray}
\label{18}
\check{R}(u)&=&-\lambda+uH,\qquad
H=iq\check{R}_q+iq^{-1}\check{R}_q^{-1},\nn\\
\check{R}(\bar u)&=&\check{R}(h)+u\bar H,\qquad \bar{H}=-iqh\check{R}_q
-iq^{-1}h^{-1}\check{R}_q^{-1},\nn\\
\check{R}^{\iota _1}(\bar u)&=&\check{R}^{\iota _1}(h)
+u\bar H^{\iota _1},\qquad \bar H^{\iota _1}=-iqh\check{R}_q^{\iota _1}
-iq^{-1}h^{-1}({\check{R}_q}^{\iota _1})^{-1},\nn\\
\check{R}^{\iota _2}(\bar u)&=&\check{R}^{\iota _1}(h^{-1})
+u\bar H^{\iota _2},\qquad \bar H^{\iota _2}=iqh^{-1}\check{R}_q^{\iota _1}
+iq^{-1}h(\check{R}_q^{\iota _1})^{-1},\nn\\
\check{R}^{\iota _1\iota _2}(u)&=&-\lambda-uH
\end{eqnarray}
into the expressions of the Monodromy matrix 
\begin{eqnarray}
\label{20}
M(u)=\check R_{01}^{\iota_2}(\bar{u})\check R_{12}(u)
\check R_{23}^{\iota_2}(\bar{u})\cdots
\check{R}_{01}^{\iota _1}(\bar u)\check{R}_{12}^{\iota _1\iota _2}(u)
\check{R}_{23}^{\iota _1}(\bar u)\cdots
\end{eqnarray}
and extract the linear terms in $u$ from the product.
By taking into account that
\begin{eqnarray}
\label{19}
\check{R}^{\iota _1}(h)\check{R}^{\iota _1}(h^{-1})&=&({\lambda}^2
+4{\sin^2 {\theta}})\hat{I},
\end{eqnarray}
with $\theta =-i \log h$ as an additive shift of the spectral
parameter, we finally 
obtain 
\begin{eqnarray}
\label{HF1} 
&-& i\lambda(\lambda^2+4\sin^2{\theta}) H_{j-1,j,j+1}=\nn\\
&=& (-1)^{j-1} \sum_{\sigma=\uparrow,\downarrow}\left\{
c^+_{j-1,\sigma}c_{j+1,\sigma}\left[(1-n_{j,\bar{\sigma}})f_4(q,h^{(-1)^j})+
n_{j,\bar{\sigma}}f_0(q,h^{(-1)^j})\right]\right.\nn\\
&+& \left.
c^+_{j+1,\sigma}c_{j-1,\sigma}\left[(1-n_{j,\bar{\sigma}})f_4(q,h^{-(-1)^j})+
n_{j,\bar{\sigma}}f_0(q,h^{-(-1)^j})\right]\right\}\nn\\
&-&(-1)^{j-1} \left\{ S^{\mu}_{j-1}S^{\mu}_{j+1} \left[(1-n_j)f_0(q,h)- n_j
\frac{f_4(q,h)+f_4(q,h^{-1})}{2} \right] \right.\nn\\
&-&\left. \left[S^3_{j-1}S^3_{j+1}+\frac{1}{4}
n_{j-1}n_{j+1}\right](2 n_j-1)f_6(q,h)\right\}
+(-1)^{j-1} \sum_{\sigma=\uparrow,\downarrow}\left\{
c^+_{j,\sigma}c_{j+1,\sigma} \right.\nn\\
&\cdot&\left[(1-n_{j+2,\bar{\sigma}})
f_{j(mod2)+1}(q,h)+(-1)^j
(1-n_{j-1,\bar{\sigma}})f_{j(mod2)+1}(q,h^{-1})\right]\nn\\
&+& \left.
c^+_{j+1,\sigma}c_{j,\sigma}\left[(1-n_{j-1,\bar{\sigma}})
f_{j(mod2)+1}(q,h)+(-1)^j
(1-n_{j+2,\bar{\sigma}}f_{j(mod2)+1}(q,h^{-1})\right]\right\}\nn\\
&+&\left[S^{\mu}_{2j-1}S^{\mu}_{2j} \frac{f_2(q,h)+f_2(q,h^{-1})}{2}
+ f_6(q,h)n_{2j-1}n_{2j}\right](n_{2j-2}-n_{2j+1})\nn\\
&+&2 S^a_{j-1}S^b_{j}S^c_{j+1}\hat{\epsilon}^{abc} +
(-1)^{j+1} \sum_{\{klm\}=\{j-1,j,j+1\}}S^{\mu}_k S^{\mu}_{lm}
\hat{\epsilon}^{klm} f_5(q,h)\nn\\
&+&f_3(q,h)(n_{j-1}n_j + n_j n_{j+1}) 
\end{eqnarray}
where
$\mu = 1,2$. 

In this expression for the Hamiltonian the spin variables are 
as in usual $t-J$ model
\bea
\label{sss}
S^{+}=c^{+}_{\uparrow}c_{\downarrow},\qquad S^{-}=
c^{+}_{\downarrow}c_{\uparrow},
\qquad S^3=\frac{n_{\uparrow}-n_{\downarrow}}{2},
\ena
the anisotropic antisymmetric tensors $\hat{\epsilon}^{abc},\quad
a,b,c=1,2,3$ 
\\and $\hat{\epsilon}^{klm}, \quad k,l,m,=j-1,j,j+1$ are
defined as follows
\bea
\label{eps}
\begin{array}{l}
\hat{\epsilon}^{a3b} = f_1(q^{1/2},h^2) \epsilon^{a3b} \\
\hat{\epsilon}^{3ab} = \hat{\epsilon}^{ab3} = f_1(q,h) \epsilon^{3ab}
\end{array}\qquad
\hat{\epsilon}^{klm} =  
\left\{\begin{array}{l}
-\hat{\epsilon}^{lkm}\qquad if  |k-l|=2, \\
\hat{\epsilon}^{lkm}\qquad  if  |k-l|=1
\end{array}
\right.
\ena
where $\epsilon^{abc}$ is ordinary antisymmetric tensor, and for
functions $f_r(q,h)$ we got
\begin{eqnarray}
\label{6F}
f_0(q,z)&=&2(h-h^{-1})^2, \nn\\
f_1(q,z)&=&2(h-h^{-1})(q^2-q^{-2}), \nn\\
f_3(q,z)&=&4(h^{-1}-h)(3q h +q h^{-1} - q^{-1}h -3q^{-1}h^{-1}),\nn\\
f_4(q,z)&=&2(q^{1/2}h^{-1}-q^{-1/2}h)^2-2(q^{1/2}-q^{-1/2})^2, \nn\\
f_5(q,z)&=&2(q-q^{-1})(h-h^{-1}), \nn\\
f_6(q,z)&=&2(q+q^{-1})(h-h^{-1})^2.
\end{eqnarray}

\begin{figure}[h]
  \begin{center}
    \leavevmode
\setlength{\unitlength}{0.0015cm}
\begingroup\makeatletter\ifx\SetFigFont\undefined
\def\x#1#2#3#4#5#6#7\relax{\def\x{#1#2#3#4#5#6}}%
\expandafter\x\fmtname xxxxxx\relax \def\y{splain}%
\ifx\x\y   
\gdef\SetFigFont#1#2#3{%
  \ifnum #1<17\tiny\else \ifnum #1<20\small\else
  \ifnum #1<24\normalsize\else \ifnum #1<29\large\else
  \ifnum #1<34\Large\else \ifnum #1<41\LARGE\else
     \huge\fi\fi\fi\fi\fi\fi
  \csname #3\endcsname}%
\else
\gdef\SetFigFont#1#2#3{\begingroup
  \count@#1\relax \ifnum 25<\count@\count@25\fi
  \def\x{\endgroup\@setsize\SetFigFont{#2pt}}%
  \expandafter\x
    \csname \romannumeral\the\count@ pt\expandafter\endcsname
    \csname @\romannumeral\the\count@ pt\endcsname
  \csname #3\endcsname}%
\fi
\fi\endgroup
{\renewcommand{\dashlinestretch}{30}
\begin{picture}(7954,2733)(0,-10)
\drawline(462,2304)(1362,504)(2262,2304)
        (3162,504)(4062,2304)(4962,504)
        (5862,2304)(6762,504)
\drawline(12,2304)(7212,2304)
\drawline(12,504)(7212,504)
\put(1182,54){\makebox(0,0)[lb]{\smash{{{\SetFigFont{12}{14.4}
{\rmdefault}$2j$}}}}}
\put(2982,54){\makebox(0,0)[lb]{\smash{{{\SetFigFont{12}{14.4}
{\rmdefault}$2j+2$}}}}}
\put(4782,54){\makebox(0,0)[lb]{\smash{{{\SetFigFont{12}{14.4}
{\rmdefault}$2j+4$}}}}}
\put(2082,2574){\makebox(0,0)[lb]{\smash{{{\SetFigFont{12}{14.4}
{\rmdefault}$2j+1$}}}}}
\put(3882,2574){\makebox(0,0)[lb]{\smash{{{\SetFigFont{12}{14.4}
{\rmdefault}$2j+3$}}}}}
\put(7662,2304){\makebox(0,0)[lb]{\smash{{{\SetFigFont{12}{14.4}
{\rmdefault}$s=1$}}}}}
\put(7662,504){\makebox(0,0)[lb]{\smash{{{\SetFigFont{12}{14.4}
{\rmdefault}$s=0$}}}}}
\end{picture}
} 
\end{center}
\caption{Zig-zag ladder chain}
\label{fig:1}
\end{figure}
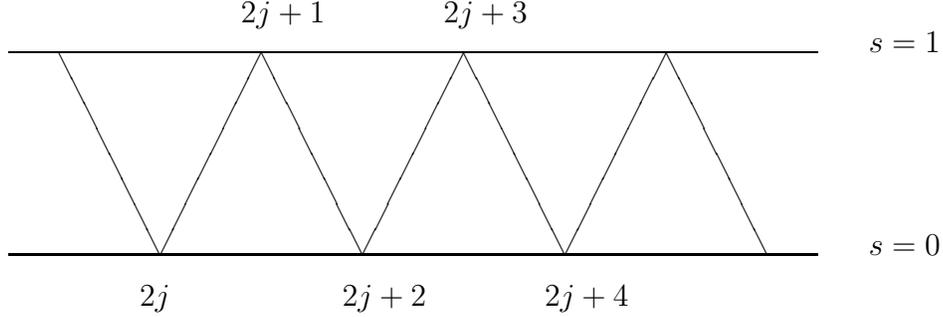

It is convenient now to write the Hamiltonian (\ref{HF1}) in a ladder form
represented graphically as in Figure \ref{fig:1}.
\def\comVa#1{{$V_{a,#1}$}}
\def\comVj#1{{$V_{j,#1}$}}
\def\comTa{{$T_1(\lambda,u)$}}
\def\comTb{{$T_0(\lambda,u)$}}
%

Let us consider the even ($2j$) and the odd ($2j+1$) points of the chain
as a sites ($j$) of two different chains labeled by $s = 0$ and $ 1$
correspondingly. The Fermi fields will be marked now as
\begin{equation}
\label{CCC}
c_{j,s}=c_{2j+s}, \qquad \qquad s = 0,1
\end{equation}

With these notations it is straightforward to obtain from the expression
(\ref{HF1}) the following ladder Hamiltonian
\begin{eqnarray}
\label{HF2} 
&-& i\lambda(\lambda^2+4\sin^2{\theta}) H_{j,s}=H_{j,s}^c
+H_{j,s}^r+H_{j,s}^{3S}+ H_{j,s}^{hS}\nn\\
&=& (-1)^s \sum_{\sigma=\uparrow,\downarrow}\left\{
c^+_{j,s,\sigma}c_{j+1,s,\sigma}\left[(1-n_{j+s,s+1,\bar{\sigma}})
f_4(q,h^{(-1)^{s+1}})+
n_{j+s,s+1,\bar{\sigma}}f_0(q,h^{(-1)^{s+1}})\right]\right.\nn\\
&+& \left.
c^+_{j+1,s,\sigma}c_{j,s,\sigma}\left[(1-n_{j+s,s+1,\bar{\sigma}})
f_4(q,h^{-(-1)^s})+
n_{j+s,s+1,\bar{\sigma}}f_0(q,h^{-(-1)^s})\right]\right\}\nn\\
&-&(-1)^{s} \left\{S^{\mu}_{j,s}S^{\mu}_{j+1,s} \left[(1-n_{j+s,s+1})
f_0(q,h)- n_{j+s,s+1}
\frac{f_4(q,h)+f_4(q,h^{-1})}{2} \right]\right.\nn\\
&-&\left.\left[S^3_{j,s}S^3_{j+1,s}+\frac{1}{4}
n_{j,s}n_{j+1,s}\right](2 n_{j+s,s}-1)f_6(q,h)\right\}
+(-1)^{j-1} \sum_{\sigma=\uparrow,\downarrow}\left\{
c^+_{j,s,\sigma}c_{j+s,s+1,\sigma}\right.\nn\\
&\cdot&\left[(1-n_{j+1,s,\bar{\sigma}})
f_{s+1}(q,h^{(-1)^s})+(-1)^s
(1-n_{j-1+s,s+1,,\bar{\sigma}})f_{s+1}(q,h^{(-1)^{s+1}})\right]\nn\\
&+& \left.
c^+_{j+s,s+1,\sigma}c_{j,s,\sigma}\left[(1-n_{j+1,s,\bar{\sigma}})
f_{s+1}(q,h^{(-1)^{s+1}})+(-1)^s
(1-n_{j-1+s,s+1,\bar{\sigma}}f_{s+1}(q,h^{(-1)^s})\right]\right\}\nn\\
&+&\left[S^{\mu}_{j,1}S^{\mu}_{j+1,0} \frac{f_2(q,h)+f_2(q,h^{-1})}{2}
+ f_6(q,h)n_{j,1}n_{j+1,0}\right](n_{j,0}-n_{j+1,1})\nn\\
&+&2 S^a_{j,s}S^b_{j+s,s+1}S^c_{j+1,s}\hat{\epsilon}^{abc} +
(-1)^{s} \sum_{\{klm\}=\{(j,s),(j+s,s+1),(j+1,s)\}}S^{\mu}_k S^{\mu}_{lm}
\hat{\epsilon}^{klm} f_5(q,h)\nn\\
&+&f_3(q,h) n_{j,s}n_{j+s,s+1}
\end{eqnarray}

Though this expression is big but one can recognise four clear terms there.
The first one is the Hamiltonian of anisotropic $t-J$ model for the
each of chains (marked as $H_{js}^c$). The second is anisotropic $t-J$ 
term for the rungs ($H_{ja}^r$). The third term ($H_{js}^{3S}$),
which was discussed in the introduction,
is written for the each triangle of the zig-zag, represents the interaction
between chains and has a topological form. The last term 
contains spin-spin  Heisenberg interaction together with hopping of
fermions and can be thought as spin-orbit interaction.

At the end of this Section let us write the $q \rightarrow 1$ limit of this
Hamiltonian , which corresponds to isotropic case.
In this limit it is necessary to rescale
the additive spectral parameter $u$ and introduce $w=\log{z}/\log{q}$
together with  new additional model parameter $\theta=\log{z}/\log{q}$,
which should be kept finite. Then we will obtain
\bea
\label{hq}
{(-1)^{s+1} \over 2}H_{js}&=&\sum_{\sigma=\uparrow,\downarrow}
c_{j,s,\sigma}^+c_{j+1.s,\sigma} + 
c_{j+1,s,\sigma}^+c_{j.s,\sigma}\nn\\
&+&\left[S_{j,s}^{\mu}S_{j+1,s}^{\mu}+S_{j,s}^{3}S_{j+1,s}^{3}
+\frac{1}{4}n_{j,s}n_{j+1,s}\right](2n_{j+s,s+1}-1)\nn\\
&-&n_{j,s}n_{j+s,s+1}n_{j+1.s} - (-1)^s 5 n_{j,s}n_{j+s,s+1},
\ena
which looks like an ordinary supersymmetric $t-J$ model
for each of chains. But the presence of the factor $2 n-1$
in front of spin-spin interaction means that it is ferromagnetic
or antiferromagnetic regarding whether the corresponding site
in the other chain is occupied by fermion or not.

\section{$ABA$ solution of the model}
\indent

The $ABA$ solution for
the homogeneous Perk-Schultz model was carried out in \cite{dV}
and in a chain of articles \cite{FK} for a model with open
boundaries.
In this section we will use the technique of Algebraic Bethe Ansatz 
(\cite{Bax}-\cite{FT}) in order to find the eigenstates and
eigenvalues of the Hamiltonian (\ref{HF1})
for staggered inhomogeneity.

At the beginning we need the definition of the $L$-operator
\begin{eqnarray}
\label{22}
L_{i,j}(z)_{m}^{m'}=\langle{m}\mid{R}_{i,j}(z)\mid{m'}\rangle=
(-1)^{p(m')p(n')}(R_{ij}(z))_{mn}^{m'n'}X_{n'}^n
\end{eqnarray}
which is a $3\times 3$ matrix in a horizontal auxiliary space, with the
matrix elements acting in quantum space. It has the following expression
\begin{eqnarray}
\label{23}
&&\hspace{-1cm}(L_{ij})_{ m}^{m'}(z)=\nn\\ 
\\
&&\hspace{-1cm} \left( 
\begin{array}{ccc}
\begin{array}{l}(qz-q^{-1}z^{-1})(1-n_{\uparrow})n_{\downarrow}\\
+(z-z^{-1})(1-n_{\downarrow})\\
\\
\end{array}&
\hspace{-7mm}
{\lambda}z^{-1}c_{\downarrow}^{+}c_{\uparrow}
\hspace{-7mm}
&  
 -\lambda z^{-1}(1-n_{\uparrow})c_{\downarrow}^{+}\\
 \lambda z c_{\uparrow}^{+}c_{\downarrow}&
\hspace{-7mm}
\begin{array}{l}
(z-z^{-1})(1-n_{\uparrow})
n_{\downarrow}\\
-(-qz+q^{-1}z^{-1})n_{\uparrow}(1-n_{\downarrow})\\
+(z-z^{-1})(1-n_{\uparrow})(1-n_{\downarrow})
\end{array}
\hspace{-7mm}
& 
-\lambda z^{-1}c_{\uparrow}^{+}(1-n_{\downarrow})\\
\\
 -\lambda z(1-n_{\uparrow})c_{\downarrow}&
\hspace{-7mm}
 -\lambda zc_{\uparrow}(1-n_{\downarrow})
\hspace{-7mm}
&
\begin{array}{l}
(z-z^{-1})(1-n_{\uparrow})n_{\downarrow}\\
+(z-z^{-1})n_{\uparrow}(1-n_{\downarrow})+\\
(q^{-1}z-qz^{-1})(1-n_{\uparrow})(1-n_{\downarrow})
\end{array}
\end{array}\right).\nn
\end{eqnarray}
The matrix elements of the monodromy operator (\ref{1}) between auxiliary
states $\mid k'\rangle$ and $\langle k\mid$ can be expressed as a
product of $L_{ij}$ matrices
\begin{eqnarray}
\label{25}
M_{0}(z)_{k'}^{k}=\langle{k}\mid{M}_{0}(z)\mid{k'}\rangle
={L}_{01}^{\iota_2}(\bar z)_{k_1}^k
L_{02}(z)_{k_2}^{k_1}\cdots L_{0N}(z)_{k'}^{k_{N-1}}, \nonumber\\
M_{1}(z)_{k'}^{k}=\langle{k}\mid{M}_{1}(z)\mid{k'}\rangle
=L_{01}^{{\iota_1}{\iota_2}}(z)_{k_1}^{k}
{L}_{02}^{\iota_1}(\bar z)_{k_2}^{k_1}\cdots 
L_{0N}^{\iota_1}(\bar z)_{k'}^{k_{N-1}}
\end{eqnarray}
which in matrix form looks like

\begin{eqnarray}
\label{26}
M_{s}(z)_{k'}^{k}=\left(
\begin{array}{lll}
A_{s,11}(z)\qquad &A_{s,12}(z)\qquad &B_{s,1}(z)\\
A_{s,12}(z)\qquad &A_{s,22}(z)\qquad &B_{s,2}(z)\\
C_{s,1}(z)\qquad &C_{s,2}(z)\qquad &D_{s}(z)
\end{array}
\right),\;\;\;\;
s=0,1,
\end{eqnarray}
where $A_{s,ab}$, $B_{s,a}$, $C_{s,a}$, $D_{s}$; $(a,b=1,2)$ 
are operators in the the quantum space.

The graded property of the model is the origin of the following form
of the transfer matrix (\ref{11})
\begin{equation}
\label{27}
\tau_{s}(z)=-A_{s,11}(z)-A_{s,22}(z)+D_{s}(z),\;\;\;\;s=0,1.
\end{equation} 

Now we would like to take the empty fermionic state 
\begin{eqnarray}
\label{28}
\mid{\Omega}\rangle_{s}=\mid{0,0,...,0}\rangle_{s}=\mid{0}\rangle_{1s}
\mid{0}\rangle_{2s}\cdots \mid{0}\rangle_{Ns},\;\;\;s=0,1,
\end{eqnarray}
as a ``test'' vacuum
and let us demonstrate that the ``vacuum'' $\mid \Omega\rangle$ is
indeed eigenstate of the transfer matrix (\ref{27})
\begin{eqnarray}
\label{29}
\tau_{s}(z)\mid{\Omega}\rangle_{s}=\nu_{s}^{(0)}\mid{\Omega}\rangle_{1-s}.
\end{eqnarray}
The action of $L_{0k}(z)$ and $L_{0k}^{\iota _2}(\bar{z})$ on the 
$k$-th empty state $\mid 0\rangle _k$
is easy to calculate
\begin{eqnarray}
\label{30}
L_{0k}(z)_m ^{m'}\mid 0\rangle &=& \left(
\begin{array}{lll}
z-z^{-1} & 0 & -\lambda z^{-1}c_{\downarrow}^+\\
0 & z-z^{-1} & -\lambda z^{-1}c_{\uparrow}^+\\
0 & 0 & q^{-1}z-qz^{-1}
\end{array}
\right)\mid 0 \rangle _k,\\
L_{0k}^{\iota _2}(\bar{z})_m ^{m'}\mid 0\rangle &=& \left(
\begin{array}{lll}
-h^{-1}z+hz^{-1} & 0 & \lambda h z^{-1}c_{\downarrow}^+\\
0 & -h^{-1}z+hz^{-1} & \lambda h z^{-1}c_{\uparrow}^+\\
0 & 0 & h^{-1}q^{-1}z-hqz^{-1}
\end{array}
\right)\mid 0\rangle _{k},
\end{eqnarray}
and as we see it has an upper-triangular form. Because the upper
triangular matrices form a semigroup the action of the monodromy matrix 
$M_{0}(z)_{k'} ^k$ on $\mid\Omega\rangle _0$ also have an
upper-triangular form directly following from the expression 
(\ref{25}) 
\begin{eqnarray}
\label{32}
&&M_{0}(z)_{m} ^{m'}\mid\Omega\rangle _0=\\
&&
\left(
\begin{array}{lll} 
\begin{array}{l}(-h^{-1}z+hz^{-1})^{N\over 2}\nn\\
(z-z^{-1})^{N\over  2}
\end{array}& 0& B_{0,1}\\
\\
0& (-h^{-1}z+hz^{-1})^{N\over 2}(z-z^{-1})^{N\over  2}& B_{0,2}\\
\\
0& 0& 
\begin{array}{l}(h^{-1}q^{-1}z-hqz^{-1})^{N\over 2}\nn\\
(q^{-1}z-qz^{-1})^{N\over 2}
\end{array}
\end{array} \right)\mid \Omega\rangle _0.
\end{eqnarray}

Following the definitions (\ref{14}-\ref{15}) of the $\iota _1$ and
$\iota _2$ operations one can easily find the actions of
$L_{0k}^{\iota _1\iota _2}(z)$ and $L_{0k}^{\iota _1}(\bar z)$ on 
$\mid 0\rangle _k$ by use of formulas
\begin{eqnarray}
\label{33}
L^{\iota _1\iota _2}(z)=L(z^{-1}),\qquad 
L^{\iota _1}(\bar z)=L^{\iota _2}({\bar z}^{-1}),
\end{eqnarray}
and correspondingly the action of $M_1(z)_{k'} ^k$ on 
$\mid\Omega\rangle _1$ will be 
\begin{eqnarray}
\label{34}
&&M_{1}(z)_{m} ^{m'}\mid\Omega\rangle _1=\\
&&
\left(
\begin{array}{lll} 
\begin{array}{l}(h^{-1}z-hz^{-1})^{N\over 2}\nn\\
(-z+z^{-1})^{N\over  2}\nn
\end{array}
& 0& B_{1,1}\\
\\
0& (h^{-1}z-hz^{-1})^{N\over 2}(-z+z^{-1})^{N\over  2}& B_{1,2}\\
\\
0& 0& 
\begin{array}{l}
(hq^{-1}z^{-1}-h^{-1}qz)^{N\over 2}\nn\\
(q^{-1}z^{-1}-qz)^{N\over 2}\nn
\end{array}
\end{array} \right)\mid \Omega\rangle _1.
\end{eqnarray}

Now we see that $\mid\Omega\rangle$ is the eigenstate of $\tau (u)$
with the eigenvalue $\nu ^{(0)}(z)$
\begin{eqnarray}
\label{35}
\nu ^{(0)}(z)&=&\nu _0 ^{(0)}(z)\nu _1 ^{(0)}(z),\nn\\
\nu _0 ^{(0)}(z)&=& -2(-h^{-1}z+hz^{-1})^{N\over 2}(z-z^{-1})^{N\over2}+
(h^{-1}q^{-1}z-hqz^{-1})^{N\over 2}(q^{-1}z-qz^{-1})^{N\over 2},\nn\\
\nu _1 ^{(0)}(z)&=&-2(h^{-1}z-hz^{-1})^{N\over 2}(-z+z^{-1})^{N\over2}+
(hq^{-1}z^{-1}-h^{-1}qz)^{N\over 2}(q^{-1}z^{-1}-qz)^{N\over 2},\nn\\
\end{eqnarray}
which follows from the expressions (\ref{11}), (\ref{27}), (\ref{29}), 
(\ref{32}). Simultaneously we see, that the $B_{s,1}(z)$ and $B_{s,2}(z)$ 
$(s=0,1)$ operators are creation operators, while the operators $C_{s}(z)$
acts on $\mid\Omega\rangle$ as the annihilation operators. This
motivates us to consider the states
\begin{eqnarray}
\label{36}
\mid{v_{1},v_{2},...v_{n}}\mid{F}\rangle_{0}=F^{{a_n}...{a_1}}
B_{0,a_1}(v_1)B_{1,a_2}(v_2)\cdots 
B_{0,a_n}(v_n)\mid{\Omega}\rangle_{0},\;\;\;a_i=1,2;\;\;
\end{eqnarray}
as  $n$-particle eigenstates of the transfer matrix $\tau (u)$. 
$F^{{a_n}...{a_1}}$
is a function of the spectral parameters $v_j$, which should be found
later.

In order to proceed further  we need to rewrite the YBEs
(\ref{12.1}) and (\ref{12.2}) in terms of the matrix 
elements of the monodromy matrix
\begin{eqnarray}
\label{37}
&&\hspace{-1cm}(-1)^{p(k'')(p(m')+p(m''))}
\check{R}_{k'm'}^{km}(\bar u,v){M_1}_{m''}^{m'}(u){M_0}_{k''}^{k'}(v)\nn\\
&&\hspace{1cm}=(-1)^{p(k')(p(m)+p(m'))}
{M_1}_{m'}^{m}(v){M_0}_{k'}^{k}(u)
\check{R}_{k''m''}^{k'm'}(\bar u,v),\nonumber\\
&&\hspace{-1cm}(-1)^{p(k'')(p(m')+p(m''))}
(\check{\tilde R}^{\iota_1})_{k'm'}^{km}(\bar u,v)
{M_0}_{m''}^{m'}(u){M_1}_{k''}^{k'}(v)\nn\\
&&\hspace{1cm}=(-1)^{p(k')(p(m)+p(m'))}
{M_0}_{m'}^{m}(v){M_1}_{k'}^{k}(u)(\check{\tilde R}^{\iota
  _1})_{k''m''}^{k'm'}(\bar u,v).
\end{eqnarray}

For our purpose, the interesting components of the equations (\ref{37}) 
are the following commutation relations between the operators
$A_{bc}(u)$ and $D(u)$ with the $B_c(v)$
\bea
\label{38}
A_{1,ba}(u)B_{0,c}(v)&=&
\frac{r_{bc}^{b'c'}(u,v)}{b(u,v)}B_{1,c'}(v)A_{0,b'a}(u)+
\frac{a_a (u,v)}{b(u,v)}B_{1,b}(u)A_{0,ca}(v),
\ena
\bea
\label{39}
A_{0,ba}(u)B_{1,c}(v)&=&
-\frac{r_{bc}^{b'c'}(u,v)}{b(u,v)}B_{0,c'}(v)A_{1,b'a}(u)-
\frac{a_a (u,v)}{b(u,v)}B_{0,b}(u)A_{1,ca}(v),
\end{eqnarray}
\begin{eqnarray}
\label{40}
D_{1}(u)B_{0,a}(v)&=&\frac{1}{b'(u,v)}B_{1,a}(v)D_{0}(u)-
\frac{{a'}_a (v,u)}{b'(v,u)}B_{1,a}(u)D_{0}(v),
\end{eqnarray}
\begin{eqnarray}
\label{41}
D_{0}(u)B_{1,a}(v)&=&-\frac{1}{b'(u,v)}B_{0,a}(v)D_{1}(u)+
\frac{{a'}_a (v,u)}{b'(v,u)}B_{0,a}(u)D_{1}(v),
\end{eqnarray}
\begin{eqnarray}
\label{42}
B_{1,a}(u)B_{0,b}(v)&=&r_{ab}^{b'a'}(u,v)B_{1,a'}B_{0,b'}(u),
\end{eqnarray}
\begin{eqnarray}
\label{43}
B_{0,a}(u)B_{1,b}(v)&=&r_{ab}^{b'a'}(u,v)B_{0,a'}B_{1,b'}(u),
\end{eqnarray}
where we have defined $b(u,v),\; a_a(u,v),\; b^{\prime}(u,v),\;
a^{\prime}_a (v,u)$ as follows
\begin{eqnarray}
\label{44}
b(u,v)&=&z(u,v)-z^{-1}(u,v),\nn\\
\nn\\
a_a(u,v)&=& \left\{ 
\begin{array}{l}
-\lambda z(u,v),\quad for \quad a=1,\\
-\lambda z^{-1}(u,v),\quad for \quad a=2,\nn\\
\end{array} 
\right.\nn\\
\nn\\
b^{\prime}(u,v)&=& -\frac{z(u,v)-z^{-1}(u,v)}{q^{-1} z^{-1}(u,v)-q
  z(u,v)},
\nn\\
\nn\\
a^{\prime}_a (v,u)&=&\left\{ 
\begin{array}{l}
\frac{\lambda z(u,v)}{q^{-1}z(u,v)-qz^{-1}(u,v)},\quad for \quad a=1,\\
\\
\frac{\lambda z^{-1}(u,v)}{q^{-1}z(u,v)-qz^{-1}(u,v)},\quad for \quad a=2
\end{array}
\right. 
\end{eqnarray}
and
\begin{eqnarray}
\label{45}
&&\hspace{-5mm}r_{bc} ^{b'c'}(u,v)=\\
\nn\\
&&\hspace{-5mm}\left(
\begin{array}{cccc}
qz(u,v)-q^{-1}z^{-1}(u,v)\hspace{-2mm}&\hspace{-2mm} 0\hspace{-2mm}&\hspace{-2mm} 0\hspace{-2mm}&\hspace{-2mm} 0\\
0\hspace{-2mm}&\hspace{-2mm} z(u,v)-z^{-1}(u,v)\hspace{-2mm}&\hspace{-2mm} \lambda z^{-1}(u,v)\hspace{-2mm}&\hspace{-2mm} 0\\
0\hspace{-2mm}&\hspace{-2mm} \lambda z(u,v)\hspace{-2mm}&\hspace{-2mm} z(u,v)-z^{-1}(u,v)\hspace{-2mm}&\hspace{-2mm} 0\\
0\hspace{-2mm}&\hspace{-2mm} 0\hspace{-2mm}&\hspace{-2mm} 0\hspace{-2mm}&\hspace{-2mm} qz(u,v)-q^{-1}z^{-1}(u,v)
\end{array} 
\right) .\nn
\end{eqnarray}
Here we have used the notation $z(u,v)=\exp{(i[u-v])}=\frac{z(u)}
{z(v)}$. 

The first terms in the commutation relations
(\ref{38})-(\ref{41}) are so called ``wanted'' terms. As we will see
they are contributing to the eigenvalues of the transfer matrix
$\tau _s(u)$, $s=0,1$. The second terms in the (\ref{38})-(\ref{41})
are so called ``unwanted'' terms and their contribution should be
canceled in order to have an eigenstate. 

Using (\ref{38})-(\ref{43}) we can represent the action of
the diagonal elements of the monodromy matrix as follows
\begin{eqnarray}
\label{46}
&&D_{1}(u)\mid{v_{1},...,v_{n}}\mid{F}\rangle_{1}=\nn\\
&&(-1)^{n\over 2}\prod_{i=1}^n
{\frac{1}{b'(\bar u,v_i)}}
[q^{-1}z^{-1}(u)-qz(u)]^{N\over 2}
[hq^{-1}z^{-1}(u)-h^{-1}qz(u)]^{N\over 2}
\mid{v_{1},...,v_{n}}\mid{F}\rangle_{0}\nn\\
&+&\sum_{k=1}^{n}
{({\tilde{\Lambda}}_{k})}_{a_{1}...a_{n}}^{b_{1}...b_{n}}
F^{a_n...a_1}
B_{1,b_{k}}(\bar u)\prod_{j=1,j\neq k}^{n}B_{b_{j}}
(v_{j})\mid{\Omega}\rangle_{1},
\end{eqnarray}
and
\begin{eqnarray}
\label{47}
&&\left [A_{1,11}(u) + A_{1,22}(u)\right] \mid v_{1},...,v_{n} \mid{F}
\rangle_{1}= \nn\\
\nn\\
&=&(-1)^{{n\over 2}+1}\prod_{i=1}^n\frac{1}{b(\bar u,v_i)}
[-z(u)+z^{-1}(u)]^{N\over 2}[h^{-1}z(u)-hz^{-1}(u)]^{N\over 2}\cdot\nn\\ 
&&{\tau^{(1)}}_{a_{1}...a_{n}}^{a'_{1}...a'_{n}}(\bar u,v_1,...,v_n)
F^{a_{n}...a_{1}}
\prod_{i=1}^{n}B_{a'_{i}}(v_{i})\mid{\Omega}\rangle_{1}+\nn\\
\nn\\
&+&\sum_{k=1}^{n}(\Lambda_{k})_{a_{1}...a_{n}}^{b_{1}...b_{n}}
F^{a_{n}...a_{1}}
B_{1,b_{k}}(\bar u)\prod_{i=1,j\neq k}^{n}B_{b_{j}}(v_{j})
\mid{\Omega}\rangle_{1},
\end{eqnarray}
where
\begin{eqnarray}
\label{48}
{\tau^{(1)}}_{a_{1}...a_{n}}^{a'_{1}...a'_{n}}(\bar u,v_1,...,v_n)=-
r_{ca_{1}}^{b_{1}a'_{1}}(\bar u,v_{1})r_{b_1a_2}^{b_2a'_2}(\bar u,v_2)
\cdots r_{b_{n-1}a_{n}}^{ca'_{n}}(u,v_{n}).
\end{eqnarray}
In the equations (\ref{46}) and (\ref{47}) the first terms appeared
due to the ``wanted'' terms of the commutation relations
(\ref{38})-(\ref{41}) while the second terms  result from the
``unwanted'' terms. 

The expression (\ref{48}) for ${\tau^{(1)}}_{a_{1}...a_{n}}^{a'_{1}...a'_{n}}$
can be rewritten as
$str[l_{n}(\bar u,v_{n})l_{(n-1)}(\bar u,v_{n-1})...l_{1}(\bar u,v_{1})]$, 
with
\begin{eqnarray}
\label{49}
[l_{k}(\bar u,v_{k})]_{b_{k-1}}^{b_{k}}&=&r_{b_{k-1}a_{k}}^{b_{k}a'_{k}}
(\bar u,v_{k})\\
&=&\left(
\begin{array}{ll}
\begin{array}{l}
[qz(\bar u,v_k)-q^{-1}z^{-1}(\bar u,v_k)]e_1 ^1\\
+(z(\bar u,v_k)-z^{-1}(\bar u,v_k))e_2 ^2
\end{array}& \lambda z^{-1}(\bar u,v_k)e_1 ^2\\
\\
\lambda z^(\bar u,v_k)e_2 ^1&
\begin{array}{l} 
[qz(\bar u,v_k)-q^{-1}z^{-1}(\bar u,v_k)]e_2 ^2\\
+(z(\bar u,v_k)-z^{-1}(\bar u,v_k))e_1 ^1 
\end{array}
\end{array}
\right).\nn
\end{eqnarray}

Here the quantum operators $e_a ^b$ act on the $k$-th space and have
$(e_a ^b)_{\beta} ^{\alpha}={\delta _a}^\alpha {\delta _\beta}^b$
matrix representation.

As it follows from the equations(\ref{46}),(\ref{47}) and
(\ref{27}), in order 
$\mid v_1,...,v_n\mid
F\rangle$ to be an eigenstate of $\tau _s(u)$
\begin{eqnarray}
\label{50}
\tau _s(u)\mid v_1,...,v_n\mid F\rangle=\nu _s(u;v_1,...,v_n,)\mid 
v_1,...,v_n\mid F\rangle
\end{eqnarray} 
we should demand

$i)$ the cancellation of the unwanted terms
\begin{eqnarray}
\label{51}
[(\tilde{\Lambda}_{k})_{a_{1}...a_{n}}^{b_{1}...b_{n}}-
({\Lambda}_{k})_{a_{1}...a_{n}}^{b_{1}...b_{n}}]F^{a_{n}...a_{1}}=0,
\end{eqnarray} 
and

$ii)$ $F$ should be an eigenvector of the small (nested) transfer
matrix $\tau ^{(1)}(u)$

\begin{eqnarray}
\label{52}
{\tau^{(1)}}_{a_{1}...a_{n}}^{a'_{1}...a'_{n}}
(u;v_{1},...,v_{n})F^{a_{n}...a_{1}}
=\nu^{(1)}(u;v_{1},...,v_n)F^{a'_{n}...a'_{1}}.
\end{eqnarray}

Once this conditions are fulfilled, the expressions for the
eigenvalues of $\tau _s(u)$  become
\begin{eqnarray}
\label{53}
&&\nu _1(u;v_1,...,v_n)=\nn\\
&=&(-1)^{n\over 2}\prod_{i=1}^n
{\frac{1}{b'(\bar u,v_i)}}
[q^{-1}z^{-1}(u)-qz(u)]^{N\over 2}
[hq^{-1}z^{-1}(u)-h^{-1}qz(u)]^{N\over 2}\nn\\
&-&
(-1)^{n\over 2}\prod_{i=1}^n\frac{1}{b(\bar u,v_i)}
[-z(u)+z^{-1}(u)]^{N\over 2}[h^{-1}z(u)-hz^{-1}(u)]^{N\over 2}
\nu ^{(1)}(u;v_1,...,v_n),
\end{eqnarray}
with
\begin{equation}
\label{54}
\nu _0(u)=\nu _1(\bar u).
\end{equation}

Therefore, as it is obvious from the preceding analysis, in order to 
know exactly the eigenvalue we need to solve an other eigenvalue-eigenstate
problem (\ref{52}) for the transfer matrix $\tau ^{(1)}$ with the
reduced amount of degrees of freedom. That is why the all procedure is
called Nested Bethe Ansatz $(NBA)$.

In the article \cite{EK} the computation of the quantities
$\tilde\Lambda _k$ and $\Lambda _k$ for the ordinary $t-J$ model was
demonstrated and the condition of cancellation of the unwanted terms
was reduced to some equations, which are defining the first set of Bethe
equations. It is not necessary to repeat this calculations here,
but let us mention the main difference, namely
\bea
\label{AAD}
(A_{1,11}+A_{1,22})\mid\Omega\rangle =
\left[-z(u)+z^{-1}(u)\right]^{N \over 2}
\left[h^{-1} z(u)-h z^{-1}(u)\right]^{N \over 2}\mid\Omega\rangle,\nn\\
D_1 \mid\Omega\rangle =
\left[-q z(u)+q^{-1} z^{-1}(u)\right]^{N \over 2}
\left[-h^{-1} z(u) q + h z^{-1}(u) q^{-1}\right]^{N \over 2}\mid\Omega\rangle,
\ena
which leads to the condition
\begin{eqnarray}
\label{59}
&&{\tau ^{(1)}}^{b' _1...b' _n} _{b_1...b_n}(v_k;v_1,...,v_n)F^{b_n...b_1}=
-\lambda \prod _{i=1;i\neq k}^n [qz(v_k,v_i)-q^{-1}z^{-1}(v_k,v_i)]\cdot\nn\\
&&\frac{[q^{-1}z^{-1}(\bar v_k)-qz(\bar v_k)]^{N\over 2}
[hq^{-1}z^{-1}(\bar v_k)-
h^{-1}qz(\bar v_k)]^{N\over 2}}{[-z(\bar v_k)+z^{-1}(\bar
v_k)]^{N\over 2}[h^{-1}z(\bar v_k)-hz^{-1}(\bar v_k)]^{N\over 2}}
F^{b' _n...b' _1}.\nn\\
\end{eqnarray}

The next step of NABA is the diagonalization of the small transfer
matrix $\tau ^{(1)}(u)$ for a chain of length $n$ 
in order for  $F$ to become an eigenvector of it.
Following the ABA technique we write the $YBE$
\begin{eqnarray}
\label{60}
\check{r}_{k'm'}^{km}(u,v){M^{(1)}}_{m''}^{m'}(u){M^{(1)}}_{k''}^{k'}(v)=
{M^{(1)}}_{m'}^{m}(v){M^{(1)}}_{k'}^{k}(u)\check{r}_{k''m''}^{k'm'}(u,v),
\end{eqnarray}
where ${M^{(1)}}_a ^{a'}$ is a monodromy matrix for the ``nested'' problem, and
\begin{eqnarray}
\label{61}
&&\check{r}_{bc} ^{b'c'}(u,v)\nn\\
\nn\\
&=&\left(
\begin{array}{cccc}
qz(u,v)-q^{-1}z^{-1}(u,v)\hspace{-3mm}&\hspace{-3mm}
0\hspace{-3mm}&\hspace{-3mm} 0\hspace{-3mm}&\hspace{-3mm} 0\\ 
0\hspace{-3mm}&\hspace{-3mm} \lambda z(u,v)\hspace{-3mm}&\hspace{-3mm}
z(u,v)-z^{-1}(u,v)\hspace{-3mm}&\hspace{-3mm} 0\\ 
0\hspace{-3mm}&\hspace{-3mm}
z(u,v)-z^{-1}(u,v)\hspace{-3mm}&\hspace{-3mm} \lambda
z^{-1}(u,v)\hspace{-3mm}&\hspace{-3mm} 0\\ 
0\hspace{-3mm}&\hspace{-3mm} 0\hspace{-3mm}&\hspace{-3mm}
0\hspace{-3mm}&\hspace{-3mm} qz(u,v)-q^{-1}z^{-1}(u,v) 
\end{array} 
\right) .\nn\\
\nn\\
\end{eqnarray}
is the $r$-matrix from (\ref{45}) in a braid formalism.

Let us now take
\begin{eqnarray}
\label{62}
M^{(1)}(u)=\left(
\begin{array}{ll}
A^{(1)}(u)&B^{(1)}(u)\\
C^{(1)}(u)&D^{(1)}(u)
\end{array}
\right),
\end{eqnarray}
and the corresponding trace 
\begin{eqnarray}
\label{63}
\tau^{(1)}(u)=-A^{(1)}(u)-D^{(1)}(u).
\end{eqnarray}
{}From the formulas (\ref{60}) one can choose following algebraic
relations
\begin{eqnarray}
\label{64}
A^{(1)}(u)B^{(1)}(v)&=&\frac{qz(u,v)-q^{-1}z^{-1}(u,v)}{z(u,v)-z^{-1}(u,v)}
B^{(1)}(v)A^{(1)}(u)\nn\\&& 
-\frac{\lambda 
z(u,v)}{z(u,v)-z^{-1}(u,v)}B^{(1)}(u)A^{(1)}(v),\nn\\
D^{(1)}(u)B^{(1)}(v)&=&\frac{q^{-1}z(u,v)-qz^{-1}(u,v)}{z(u,v)-z^{-1}(u,v)}
B^{(1)}(v)D^{(1)}(u)\nn\\&& 
+\frac{\lambda z(u,v)}{z(u,v)-z^{-1}(u,v)}
B^{(1)}(u)D^{(1)}(v),\nn\\
B^{(1)}(u)B^{(1)}(v)&=&B^{(1)}(v)B^{(1)}(u),
\end{eqnarray}
which are the $YBE$ for the $XXZ$ model with staggered inhomogeneity,
defined in the article \cite{APSS}.
The first (second) term in the expression (\ref{64}) 
called ``wanted'' (correspondingly
``unwanted'') term.

Let us take
\begin{eqnarray}
\label{65}
\mid 0\rangle_{k}^{(1)} =\left(
\begin{array}{l}
1\\
0
\end{array}
\right),\;\;\nn\\
\mid\Omega\rangle^{(1)}=\mid 0\rangle_1 ^{(1)}
\dots\mid 0\rangle_n ^{(1)}=\bigotimes_{k=1}^n\mid 0\rangle_k ^{(1)}.
\end{eqnarray}
as a reference state.
In order to find the action of nested monodromy matrix $M^{(1)}(u)$ on
the reference state one should act by $l_k(u)$ from (\ref{49}) on 
$\mid 0\rangle _k ^{(1)}$.

After simple calculations we obtain

\begin{eqnarray}
\label{66}
A^{(1)}(u)\mid \Omega\rangle ^{(1)}=\prod _{i=1} ^{n}[qz(u,v_i)
-q^{-1}z^{-1}(u,v_i)]\mid\Omega\rangle ^{(1)},\nn\\
D^{(1)}(u)\mid\Omega\rangle ^{(1)}=\prod _{i=1}
^{n}[z(u,v_i)-z^{-1}(u,v_i)]\mid\Omega\rangle ^{(1)}.
\end{eqnarray}

Now let us make the following Ansatz for the eigenstates of $\tau ^{(1)}(u)$ 

\begin{eqnarray}
\label{67}
\mid v_{1}^{(1)},\dots,v_m^{(1)}\rangle=B^{(1)}(v_1 ^{(1)})B^{(1)}(v_2 ^{(1)})
\dots B^{(1)}(v_m ^{(1)})\mid\Omega\rangle^{(1)}.
\end{eqnarray}
By use of (\ref{64}) the actions of $A^{(1)}(u)$ and $D^{(1)}(u)$ 
on the states (\ref{67}) is given by
\begin{eqnarray}
\label{68}
 D^{(1)}(u)\mid v_1^{(1)},\dots ,v_m^{(1)}\rangle \hspace{-3cm}&&
\nn\\
&=&\prod_{j=1}^m
\frac{q^{-1}z(u,v_i ^{(1)})-qz^{-1}(u,v_i ^{(1)})}{z(u,v_i ^{(1)})-
z^{-1}(u,v_i ^{(1)})}\prod _{j=1} ^{n}[z(u,v_j)-z^{-1}(u,v_j)]
\mid v_1^{(1)},\dots ,v_m^{(1)}\rangle \nn\\
&&+\sum_{k=1}^m\Lambda_k^{(1)}B^{(1)}(u)
\prod_{j=1,j\neq k}^m B^{(1)}(v_j ^{(1)})\mid\Omega\rangle^{(1)},\nn\\
A^{(1)}(u)\mid v_1^{(1)},\dots ,v_m^{(1)}\rangle \hspace{-3cm}&&
\nn\\
&=&\prod_{i=1}^m
\frac{qz(u,v_i^{(1)})-q^{-1}z^{-1}(u,v_i^{(1)})}{z(u,v_i^{(1)})-
z^{-1}(u,v_i^{(1)})}\prod_{j=1}^n[qz(u,v_i)-q^{-1}z^{-1}(u,v_i)]
\mid v_1^{(1)},\dots ,v_m^{(1)}\rangle\nn\\
&&+\sum_{k=1}^m\tilde{\Lambda}_k^{(1)}B^{(1)}(u)\prod_{j=1,j\neq k}^m 
B^{(1)}(v_j ^{(1)}) \mid\Omega\rangle^{(1)}.
\end{eqnarray}

The first terms in this expressions are  the ``wanted'' terms,
while the second terms (with $\Lambda _k ^{(1)}$, $\tilde{\Lambda}_k
^{(1)}$) are the ``unwanted'' terms of the nested problem 
and should cancel each other
in the eigenvalue equation (\ref{52}). Hence we can write the
eigenvalues $\nu ^{(1)(u;v_1,\dots,v_n;v^{(1)} _1,\dots ,v^{(1)}_m)}$
of $\tau ^{(1)}(u)$ 
\begin{eqnarray}
\label{69}
\tau ^{(1)}(u)\mid v_1^{(1)},\dots ,v_m^{(1)}\rangle=
[-A^{(1)}(u)-D^{(1)}(u)]\mid v_1^{(1)},\dots ,v_m^{(1)}\rangle\nn\\
=
-\left [\prod_{i=1}^m
\frac{qz(u,v_i^{(1)})-q^{-1}z^{-1}(u,v_i^{(1)})}{z(u,v_i^{(1)})-
z^{-1}(u,v_i^{(1)})}\prod
_{j=1}^n[qz(u,v_i)-q^{-1}z^{-1}(u,v_i)]\right.
\nn\\
\left. +\prod _{j=1}^m
\frac{q^{-1}z(u,v_i ^{(1)})-qz^{-1}(u,v_i ^{(1)})}{z(u,v_i ^{(1)})-
z^{-1}(u,v_i ^{(1)})}\prod _{j=1} ^{n}[z(u,v_j)-z^{-1}(u,v_j)]\right]
\mid v_1^{(1)},\dots ,v_m^{(1)}\rangle.
\end{eqnarray} 
Setting now $u=v_k$ in (\ref{69}) and comparing it with the
condition (\ref{59}) we obtain the first set of Bethe equations
\begin{eqnarray}
\label{70}
\prod_{i=1}^m
\frac{qz(v_k,v_i^{(1)})-q^{-1}z^{-1}(v_k,v_i^{(1)})}{z(v_k,v_i^{(1)})-
z^{-1}(v_k,v_i^{(1)})}
&=&\frac{[h^{-1}q^{-1}z(v_k)-hqz^{-1}(v_k)]^{N\over 2}[q^{-1}z(v_k)-
qz^{-1}(v_k)]^{N\over 2}}{[-h^{-1}z(v_k)+hz^{-1}(v_k)]^{N\over 2}
[z(v_k)-z^{-1}(v_k)]^{N\over2}},
\nn\\
k&=&1,\dots,n .
\end{eqnarray}

The equations which are obtained from the cancellation of the unwanted terms 
$\Lambda ^{(1)} _k$ and $\tilde{\Lambda} ^{(1)} _k$  
\begin{eqnarray}
\label{71}
\prod _{i=1,i\neq k}^m\frac{q^{-1}z(v_k ^{(1)},v_i ^{(1)})-qz^{-1}
(v_k ^{(1)},v_i ^{(1)})}
{qz(v_k ^{(1)},v_i ^{(1)})-q^{-1}z^{-1}(v_k ^{(1)},v_i ^{(1)})}=
\prod _{j=1}^n\frac{qz(v_k ^{(1)},v_j)-q^{-1}z^{-1}(v_k ^{(1)},v_j)}
{z(v_k ^{(1)},v_j)-z^{-1}(v_k ^{(1)},v_j)}, \nn\\ 
k=1.\dots ,m
\end{eqnarray}
are the second set of Bethe equations.

Finally we obtain the eigenvalues of $\tau _1(u)$ and $\tau _0(u)$
from (\ref{53}) and (\ref{54}) as
\begin{eqnarray}
\label{72}
\nu _1(u;\{v\};\{v^{(1)}\})&=&(-1)^{n\over 2}\prod_{i=1}^n
{\frac{1}{b'(\bar u,v_i)}}
[q^{-1}z^{-1}(u)-qz(u)]^{N\over 2}
[hq^{-1}z^{-1}(u)-h^{-1}qz(u)]^{N\over 2}\nn\\
&&+
(-1)^{n\over 2}\prod_{i=1}^n\frac{1}{b(\bar u,v_i)}
[-z(u)+z^{-1}(u)]^{N\over 2}[h^{-1}z(u)-hz^{-1}(u)]^{N\over 2}\cdot
\nn\\
&&\cdot\left [\prod_{i=1}^m
\frac{qz(u,v_i^{(1)})-q^{-1}z^{-1}(u,v_i^{(1)})}{z(u,v_i^{(1)})-
z^{-1}(u,v_i^{(1)})}\prod
_{j=1}^n[qz(u,v_i)-q^{-1}z^{-1}(u,v_i)]\right.
\nn\\
&&\left. +\prod _{j=1}^m
\frac{q^{-1}z(u,v_i ^{(1)})-qz^{-1}(u,v_i ^{(1)})}{z(u,v_i ^{(1)})-
z^{-1}(u,v_i ^{(1)})}\prod _{j=1} ^{n}[z(u,v_j)-z^{-1}(u,v_j)]\right]
,\nn\\
\end{eqnarray} 
and
\begin{eqnarray}
\label{73}
\nu _0(u;\{v\};\{v^{(1)}\})=\nu _1(\bar u;\{v\};\{v^{(1)}\}).
\end{eqnarray}

\section{Acknowledgements}
\indent

The author acknowledge D.Arnaudon, A.Sedrakyan, P.Sorba
for the formulation of the problem and many fruitful
discussions.

Work was partially supported by INTAS
grant 99-1459.


\end{document}